\documentclass{article}
\usepackage{spconf,amsmath,graphicx}
\usepackage{multirow}
\usepackage[hang,flushmargin]{footmisc}
\usepackage[textfont=it,tableposition=top]{caption}
\usepackage{hyperref}
\hypersetup{colorlinks=true}

\def \cleanCgm          {\mathbf X}
\def \EstCleanCgm       {\widehat{\mathbf X}}
\def \noiseCgm          {\mathbf N}

\def \mixCgm            {{\mathbf Y}}   %

\def \IRM               {M}           %
\def \ERM               {\widehat{M}}   %
\def \TI                {t}             %
\def \FI                {c}

\title{Cross-attention conformer for context modeling in speech enhancement for ASR}
\name{Arun Narayanan, Chung-Cheng Chiu, Tom O'Malley, Quan Wang, Yanzhang He}
\address{Google LLC, U.S.A.}
\begin{document}
\maketitle
\begin{abstract}
This work introduces \emph{cross-attention conformer}, an attention-based architecture for context modeling in speech enhancement. Given that the context information can often be sequential, and of different length as the audio that is to be enhanced, we make use of cross-attention to summarize and merge contextual information with input features. Building upon the recently proposed conformer model that uses self attention layers as building blocks, the proposed cross-attention conformer can be used to build deep contextual models. As a concrete example, we show how noise context, i.e., short noise-only audio segment preceding an utterance, can be used to build a speech enhancement feature frontend using cross-attention conformer layers for improving noise robustness of automatic speech recognition. 
\end{abstract}
\begin{keywords}
Noise robust ASR, noise context, speech separation, ideal ratio mask
\end{keywords}
\section{Introduction}
\label{sec:intro}

Performance of automatic speech recognition (ASR) systems in the presence of noise has improved quite significantly over the last few years. The use of neural network based acoustic models ~\cite{PrabhavalkarRaoSainathLiEtAl17,BattenbergChenChildCoatesEtAl17,HoriWatanabeZhangChan2017,li2021betterfaster}, and large scale training ~\cite{mirsamadi2017multi,hakkani2016multi,NarayananMisraSimPundakEtAl18} have been the main contributing factors. Augmenting training data using simulated noisy utterances ~\cite{kim2017mtr}, and other strategies \cite{park2019specaugment, medennikov2018investigation} have also played important roles. Even with these improvements, performance of ASR models still degrades in the presence of significant background noise \cite{barker2017thirdchime,barker2018fifthchime}. Speech enhancement frontends for ASR that specifically address background noise have therefore been widely studied \cite{Li2014Review}. 

Typical speech enhancement frontends map noisy speech to clean speech either in the waveform domain \cite{chen2021continuous,luo2018tasnet} or in the ASR feature domain \cite{Narayanan2013IRM}. But, in a number of applications, it is possible to have additional contextual signals that can aid enhancement. Contextual signals provide useful information about the acoustic scene in which the noisy audio is recorded. Such information cannot be easily inferred just from the segment of audio that is to be enhanced. Examples include noise context, i.e., a short segment of noise preceding speech, and speaker embedding or speaker enrollment utterance \cite{wang2021tunein}. A notable example of an enhancement model that makes use of contextual information is the SEANet \cite{tagliasacchi2020seanet}, which uses synchronized accelerometer data as contextual input to perform enhancement. 

Noise context has been used in prior works for improving robustness of ASR, and  wake-word detection models. In noise-aware training  \cite{Seltzer2013DNNAurora4}, features from the first few noise-only frames are averaged and stacked with input features. This approach works mainly for stationary noise types, since the averaged feature does not capture variations. Noise context is successfully used in the Multichannel Cleaner algorithm \cite{huang2019hotwordcleaner} for learning a noise-cancellation filter. The approach is multi-channel, and makes use of the strong correlation between noise signals recorded at multiple microphones to learn the filters. The filters are kept fixed and used to remove noise from the audio containing target speech or the wake-word. Motivated by this, in the current work, we make use of the noise context but build a single-channel enhancement frontend using neural networks. Even without using multiple microphone channels, our results show that the spectral characteristics derived from the noise context can provide useful information for improving performance.

When using contextual signals as additional input, typically, they are assumed to be either in sync or summarized to a single embedding. For example, in noise-aware-training, the noise is averaged to a vector or a noise embedding; speaker enrollment audio is typically converted to a speaker embedding (a.k.a. d-vector) \cite{wang2019voicefilter,Wang2020}. The summarized information can then be easily appended to each frame of the noisy input audio. Other contextual signals, like accelerometer data in SEANet, or synchronized noise estimate \cite{Narayanan2015djat}, are of the same length as the input audio, making it easy to stack them with the noisy input. But, in reality, not all contextual signals can be treated in these ways without loss of information. Noise context signals, for example, can be of different lengths than the input audio, and summarizing it into a single embedding will likely result in loss of information. Similarly, a device can have multiple speaker enrollments, resulting in multiple d-vectors -- one per enrolled speaker -- and it may not be known, \textit{a priori}, which speaker is active. In this work, relying on the cross-attention mechanism \cite{vaswani2017attention}, we develop a model architecture that makes it easy to incorporate contextual signals that are not necessarily of the same length as the input or be easily summarized into a fixed length embedding.

Attention-based sequence-to-sequence models are now the leading architecture for building speech applications \cite{gulati2020conformer,yeh2019transformer,karita2019comparative,subakan2021attention,chen2021continuous}. Most models use multiple layers of self-attention blocks to encode acoustic features into a hidden representation. For ASR, the encoded features are then decoded to word sequences using recurrent transducers \cite{graves12rnnt} or attention-based models \cite{chan2016listen}. Attention-based decoders are LSTM-based \cite{chan2016listen}, or themselves use blocks of self-attention and cross-attention \cite{li2020parallel}. Cross-attention allows the decoder layers to attend to encoded acoustic features when predicting the sequence of word tokens. While prior work on cross-attention-based models have mainly used the standard transformer architecture \cite{li2020parallel,ding2020textual}, the recently proposed conformer architecture \cite{gulati2020conformer} for self-attention has been shown to work well when processing acoustic features. In this work, we propose extending the conformer architecture to build cross-attention blocks for incorporating contextual information.

Using noise-context modeling as an example of context modeling in speech enhancement, we show that conformer-based cross-attention blocks can be used to obtain significant improvements in performance in noisy conditions, in terms of word error rate (WER). When using a large-scale state-of-the-art ASR system \cite{sainath2020streaming}, using noise context improves WERs by 9\% -- 28\% relative to the noisy baseline, and 5\% -- 19\%, relative to a model that does not use noise context. Since this work specifically focuses on ASR, the enhancement frontend operates directly in the feature domain without the need to reconstruct audio in the waveform space. But the presented models can be readily used for speech enhancement in the waveform domain for improving speech quality.

The rest of the paper is organized as follows. In {Section~\ref{sec:model}} we present the cross-attention conformer and various modeling choices when constructing the layer. {Section~\ref{sec:expr}} presents our experimental setup, which is followed by results in {Section~\ref{sec:results}}. Conclusions and a discussion of future work are presented in {Section~\ref{sec:concl}}.

\section{Cross-attention Conformer}
\label{sec:model}

\begin{figure}[h]
  \centering
  \includegraphics[scale=0.6]{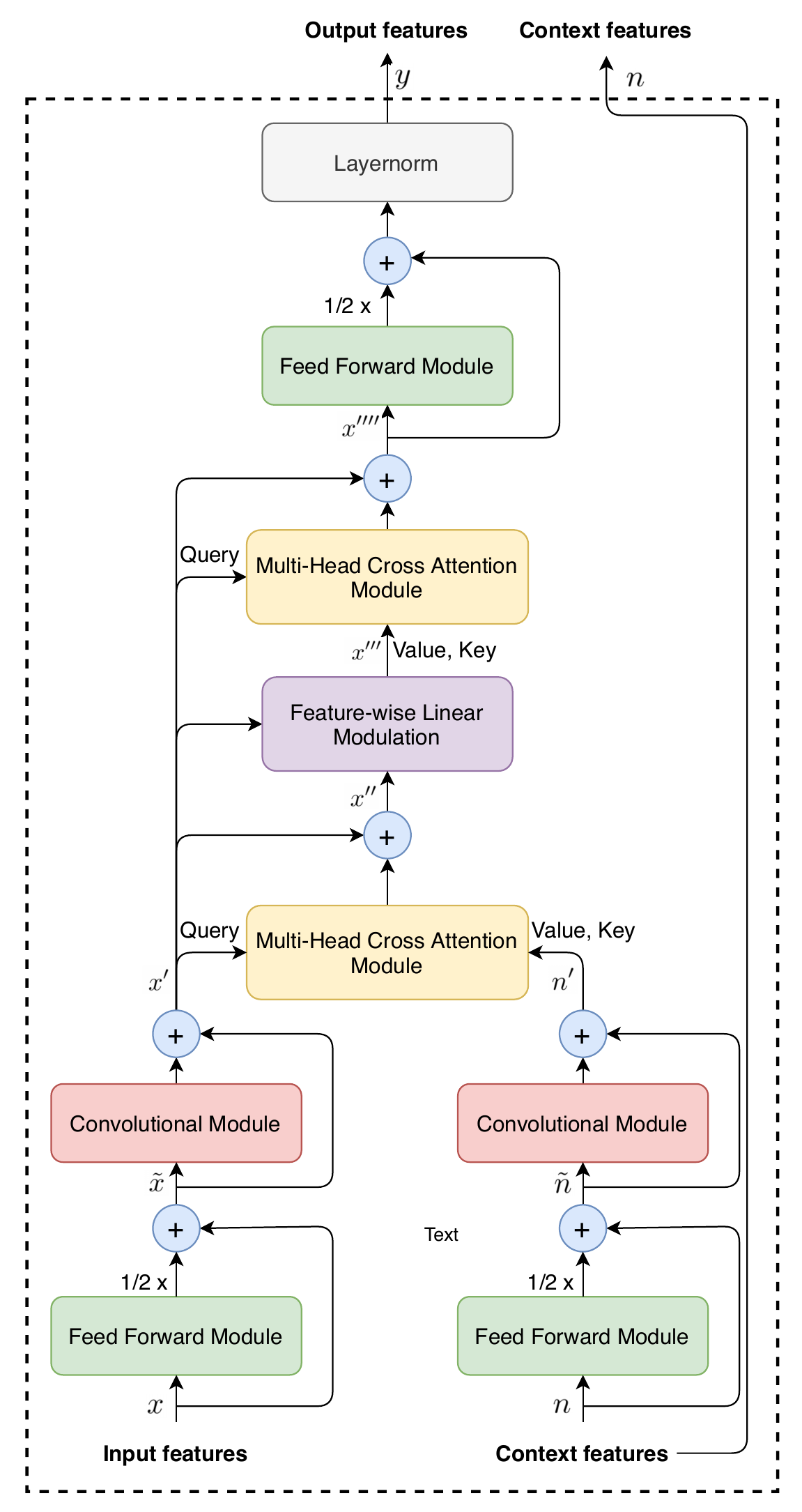}
  \caption{{The proposed cross-attention conformer layer that extends the conformer layer to make use auxiliary inputs.}}
   \label{fig:ca-conformer}
   \vspace{-0.1in}
\end{figure}

A block diagram of the proposed cross-attention conformer is shown in Fig.~\ref{fig:ca-conformer}. Similar to a conformer layer, which is shown in Fig.~\ref{fig:conformer}, the cross-attention conformer layer consists of blocks of feed-forward modules and convolutional modules. Motivated by the use of cross-attention for summarizing speaker information in \cite{wang2021tunein}, the proposed model replaces multi-headed self-attention with multi-headed cross-attention modules, and uses an additional feature-wise linear modulation layer \cite{perez2018film} to merge context features with input features. These modules are described in detail in the following subsections.

\subsection{Conformer}

\begin{figure}[h]
  \centering
  \includegraphics[scale=0.7]{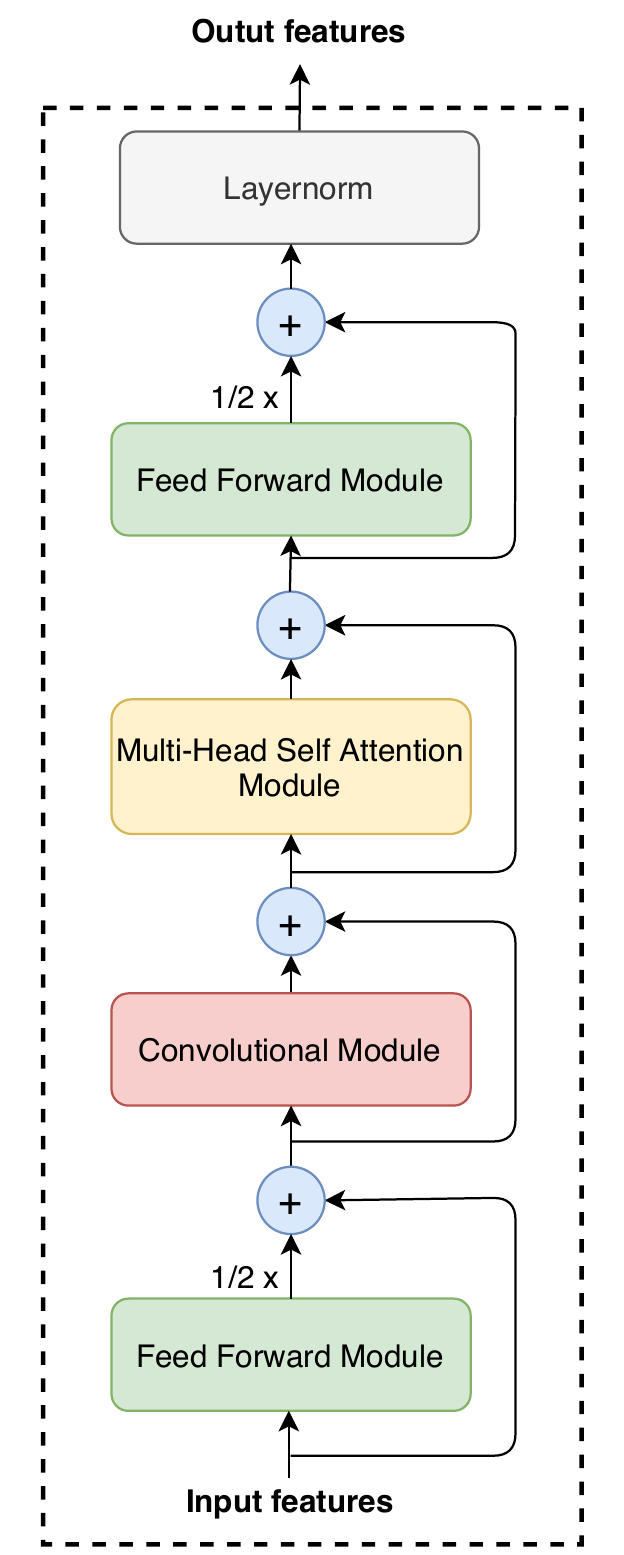}
  \caption{{Conformer architecture that uses convolution and self-attention.}}
   \label{fig:conformer}
   \vspace{-0.1in}
\end{figure}

As shown in Fig.~\ref{fig:conformer}, a conformer layer consists of a sequence of a half-step feed-forward module, a convolutional module, a self-attention module, and a final half-step feed-froward module. The architecture combines convolution and self-attention to capture short-term and longer term interactions in the input features, respectively. Overall, we very closely follow the architecture introduced in \cite{gulati2020conformer}, with minor modifications outlined in \cite{li2021betterfaster}. Specifically, unlike \cite{gulati2020conformer}, the convolutional module appears before the self-attention module, which enables the self-attention blocks to not use relative positional embedding \cite{DaiYangYangCarbonnelEtAl19}, since it can be partly captured by temporal convolution \cite{wang2020transformer}. The attention block uses multi-headed self-attention (without relative positional embedding). The convolutional block uses point-wise convolution, gated linear units, 1-D depth-wise convolution, and group normalization. There are residual connections between each block, and layer normalization before each processing block, and after the final half-step feed-forward module.

To build a conformer-based model, the conformer block can be stacked to extract an encoded feature representation, which can then be used to estimate the target of interest.

\subsection{Cross-attention Conformer block}
\label{subsec:caconformer}

Apart from the input features, the cross-attention conformer also receives context features as auxiliary input. Similar to the first few modules of a conformer, both input features and auxiliary features are first processed through half-step feed-forward modules, and convolutional modules. This is followed by a multi-headed cross-attention module that uses the processed input features to derive the \emph{query} vector, and the processed auxiliary features to derive the \emph{key} and \emph{value} vectors \cite{bahdanau2014neural,vaswani2017attention}. The goal of this first cross-attention block is to summarize the context features independently for each frame of the processed input feature. This allows the model to learn a weighted combination of the frames of the noise context features for each input frame to be enhanced. The summarized noise context features are essentially input-frame-specific noise embeddings. Unlike prior work that computes a single noise embedding to be used by every input frame \cite{Seltzer2013DNNAurora4}, computing input-frame-specific noise embeddings allow the model to better handle non-stationary noise types that have a lot of temporal variations and cannot be easily summarized using a single embedding vector.

The first cross-attention module is followed by a merge operation, that combines the context features with the input features using feature-wise linear modulation (FiLM) \cite{perez2018film}:
\begin{equation}
    \textrm{FiLM}(x, y) = r(y) \odot x + h(y).
\end{equation}
Here, $r$ and $h$ are affine transforms, and $x$ and $y$ are the inputs to the FiLM layer. $\odot$ stands for pointwise multiplication. FiLM has been shown in prior work to be more expressive when combining multiple feature representations as opposed to the simpler weighted combination. 

The FiLM layer is followed by another cross-attention layer that merges the input features, which are again used as query vectors, and the output of the FiLM layer, which is used to derive the key and value vectors. The goal of the second cross-attention is to learn longer-term interactions in the processed feature, similar to the self-attention block in a conformer. 

In Sec.~\ref{sec:results}, we present ablations that show the benefit of the FiLM module and the second cross-attention module.

Similar to the conformer, each processing module starts with a layer norm; there are residual connections between the modules. When not using FiLM and a second cross-attention layer, residual connections act as a simple mechanism to combine the summarized context features from the output of the first cross-attention module with the input features.

Mathematically, a cross-attention conformer layer transforms its input features, $x$, and context features, $n$, to output features, $y$, as follows:
\begin{align}
\begin{split}
\tilde{x} &= x + \frac{1}{2}\textrm{FFN}(x), \tilde{n} = n + \frac{1}{2}\textrm{FFN}(n),  \\
x^{\prime} &= \tilde{x} + \textrm{Conv}(\tilde{x}), n^{\prime} = \tilde{n} + \textrm{Conv}(\tilde{n}), \\
x^{\prime\prime} &= x^{\prime} + \textrm{MHCA}(x^{\prime}, n^{\prime}), \\
x^{\prime\prime\prime} &= x^{\prime} \odot r(x^{\prime\prime}) + h(x^{\prime\prime}), \\
x^{\prime\prime\prime\prime} &= x^{\prime} + \textrm{MHCA}(x^{\prime}, x^{\prime\prime\prime}), \\
y &= \textrm{LayerNorm}(x^{\prime\prime\prime\prime} + \frac{1}{2} \textrm{FFN}(x^{\prime\prime\prime\prime})). \\
\end{split}
\end{align}
Here, FFN, Conv, and MHCA, stand for feed-forward module, convolution module and multi-headed cross-attention module, respectively.

Cross-attention conformer blocks can be stacked to learn deeper representations. In practice, as we describe in the following section, we typically use conformer blocks that independently process input features and context features; the cross-attention layers operate on the encoded features and the encoded context. When stacking multiple cross-attention layers, the encoded context is passed as auxiliary input to each of those layers.
\section{Experimental settings}
\label{sec:expr}

\subsection{Data}
All of our models are trained on a combination of $281$k utterances from the Librispeech training set \cite{panayotov2015librispeech} and $1,916$k utterances from an internal, vendor collected dataset. The noisy utterances are generated using a room simulator \cite{kim2017mtr}, with signal-to-noise (SNR) ratio ranging from {$-10$~dB} to {$30$~dB}. Noise is sampled from internally collected noise snippets that simulate conditions like cafe, kitchen, and cars, and freely available noise sources from Getty\footnote{\url{https://www.gettyimages.com/about-music}} and YouTube Audio Library\footnote{\url{https://youtube.com/audiolibrary}}. Room configurations have reverberation times ($T60$) ranging from 0 msec to 900 msec. We generate multiple copies of the data under different mixing conditions in order to model enough combinations of clean speech, background noise, and room-configuration.

A 6-second noise-only segment preceding a query is assumed to be available to model noise context. This is simulated during training by choosing longer segments of noise compared to target queries \footnote{Some target utterances are long, e.g., in Librispeech, making it harder to choose noise segments longer than target speech. In those cases, only some initial portion of the noisy speech will have the same acoustic characteristics as the noise segment.}.

For evaluations, we use 3 groups of noisy sets:
\vspace{-0.1in}
\begin{itemize}
    \setlength\itemsep{0.1em}
    \item The first group is created by mixing the test-clean subset of Librispeech with similar, but held-out, noise segments as the training set. 3 subsets, at SNRs {$-5$~dB},  {$0$~dB}, and  {$5$~dB}, are constructed. There are $\sim$2.6k utterances in each subset. 
    \item The second group is created using a held-out vendor-collected test set, which was re-recorded in a room using a mouth simulator. These utterances are then mixed with re-recorded noise to simulate a movie playing in the background when a query is spoken. The utterances are mixed at SNRs, {$0$~dB},  {$6$~dB}, and  {$12$~dB}. There are $\sim$2k utterances in each subset. 
    \item Finally, we also evaluate performance in multi-talker conditions. We mix Librispeech test-clean and test-other subsets with competing speech from Librispeech and vendor collected sets at SNRs, {$-5$~dB},  {$0$~dB}, and  {$5$~dB}. Although our model is not trained on this type of data, we expect the noise context to carry information about competing speech that the model can use for enhancement. Competing speech is also highly non-stationary; therefore, this set helps measure how well our models can generalize to previously unseen and challenging conditions. There are $\sim$5.5k utterances in each subset.
\end{itemize}

\subsection{Overall architecture}

\begin{figure}[ht]
  \centering
  \includegraphics[scale=0.5]{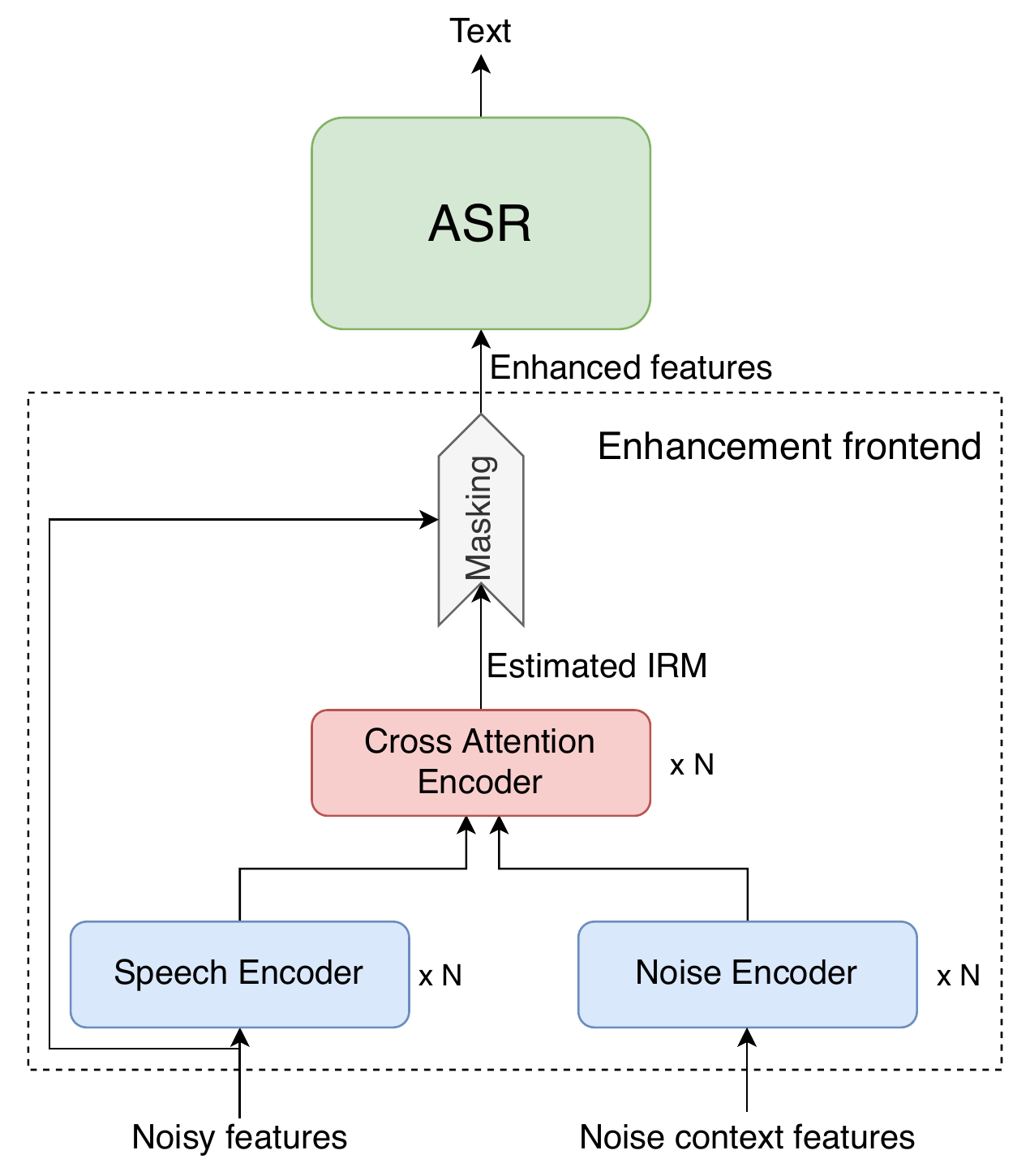}
  \caption{{A block diagram of the overall architecture. The enhancement frontend consists of speech and noise encoders, and the cross-attention encoder, each of which can have multiple layers. The ASR model takes enhanced features as input.}}
   \label{fig:model-archi}
   \vspace{-0.1in}
\end{figure}

The overall architecture of the model is shown in Fig.~\ref{fig:model-archi}. It consists of an enhancement frontend and a pretrained ASR model. The enhancement frontend takes noisy input features and noise context features, and outputs enhanced features, which are then fed to the ASR model for final transcription. The ASR model and the enhancement frontend are trained independently and on disjoint training sets. Each of these modules are described in detail the following subsections. 

\subsection{Enhancement frontend}

\subsubsection{Features}
The enhancement frontend operates in the feature space, without the need to transform enhanced audio back to the time domain. As features, we use 128-dimensional log Mel spectrograms. The window size is set to 32 msec, with a hop size of 10 msec. Log Mel spectrograms are computed for both input features and noise context using the same parameters. 

\subsubsection{Targets}
The enhancement frontend uses the ideal ratio mask (IRM) as the training target \cite{Narayanan2013IRM}. IRMs are computed using reverberant speech and reverberant noise, with the assumption that speech and noise are uncorrelated in the Mel spectral space:
\begin{equation}
\label{eq:irm}
\IRM(\TI,\FI) = \frac{\cleanCgm(\TI,\FI)}{\cleanCgm(\TI,\FI) + \noiseCgm(\TI,\FI)}.
\end{equation}
Here, $\cleanCgm$ and $\noiseCgm$ are the reverberant speech and the reverberant noise Mel spectrograms, respectively. $\TI$ and $\FI$, represent time and Mel frequency bin indices. We choose to estimate IRMs because the targets are bounded between $[0, 1]$, simplifying the estimation process. Moreover, the ASR model we use for evaluation is trained on real and simulated reverberant data, making it relatively robust to reveberant speech. Therefore, IRMs derived using reverberant speech as the target still provide substantial gains in performance.

\subsubsection{Loss}
The loss used during training is a combination of $\ell_1$ and $\ell_2$ losses between the IRM and estimated IRM, $\ERM$:
\begin{equation}
\label{eq:irm_loss}
\mathcal{L} = \sum_{\TI,\FI} \underbrace{|\IRM(\TI,\FI) - \ERM(\TI,\FI)|}_{\ell1} + \underbrace{(\IRM(\TI,\FI) - \ERM(\TI,\FI))^2}_{\ell2}.
\end{equation}

\subsubsection{Inference}
During inference, the estimated IRM is scaled and floored to reduce speech distortion at the expense of reduced noise suppression. This is especially important, since the ASR model is sensitive to speech distortions and non-linear frontend processing, which is one of the main challenges in improving performance of robust ASR models using enhancement frontends \cite{Narayanan2014Joint}. The enhanced feature is derived as:
\begin{equation}
\label{eq:masking}
\EstCleanCgm(\TI,\FI) = \mixCgm(\TI,\FI) \odot \max(\ERM(\TI,\FI), \beta)^\alpha.
\end{equation}
Here, $\mixCgm$ is the noisy Mel spectrogram, $\EstCleanCgm$ is an estimate of the clean Mel spectrogram, $\alpha$ and $\beta$ are exponential mask scalar and mask floor. In our evaluations, $\alpha$ is set $0.5$, and $\beta$ is set to $0.01$. The enhanced features are log-compressed, i.e. $log(\EstCleanCgm)$, and passed to the ASR model for evaluation.

\subsubsection{Enhancement model architecture}
The baseline enhancement frontend with no noise context consists of 4 conformer layers for the speech encoder, each with 512 units. The kernel size for depthwise convolution within the convolutional module of the conformer is set to 15. For the self-attention blocks, the model uses masked attention with 8 attention heads, with each frame attending to 64 frames in the past. We do not use full attention models so as to build streaming enhancement models with no right context. After the final conformer layer, a fully connected layer with sigmoid activation function maps the encoded feature to the output. The model has $\sim24$M parameters.

The enhancement frontend that uses noise context has 2 conformer encoder layers each to process input features and noise context independently. 2 layers of cross attention conformers then process the encoded features and context. The model dimension of the conformer layers is set to 256, so that the model has $\sim$19M parameters, which is comparable to the baseline model. The number of attention heads and the kernel size for convolution is the same as the baseline model.

\subsubsection{ASR model}
The ASR model used for evaluations is an LSTM based recurrent neural net transducer \cite{sainath2020streaming}. The model is trained on $\sim$400k hours of speech from a varied set of domains like VoiceSearch, YouTube, Telephony, and Farfield, and consists of anonymized and hand-transcribed English utterances. The model also uses data augmentation to simulate noisy conditions during training with SNRs between 0~dB and 30~dB, and reverberation times between 0~msec and 900~msec. This model is trained using log Mel spectrogram features computed the same way as those used by the enhancement frontend. Note that the ASR model is assumed to be pre-trained in this work, independent of the enhancement frontend.

All models are trained in TensorFlow \cite{abadi2016tensorflow} using the Lingvo toolkit \cite{ShenNguyenWuChenEtAl19}. We use Tensor Processing Units \cite{jouppi2017datacenter} and Adam optimization \cite{kingma2014adam}. 
\section{Results}
\label{sec:results}

\subsection{Noise-context models}

\begin{table}[tbh]
  \centering
  \caption{Results using noise context models on Librispeech test-clean subset with added reverberation and background noise. The baseline model uses no enhancement frontend. E0 corresponds to a frontend that does not use noise context. E3 is the model described in Sec.~\ref{sec:model} with 2 cross-attention and the FiLM modules. E2 is model without FiLM. E1 is a model without FiLM or the second cross-attention.}
  \label{tab:librispeech-noisecontext}
  \begin{tabular}{lcccc}
    \hline
    \multirow{2}{*}{\textbf{Test set}} & \multicolumn{3}{c}{\textbf{SNR}} & \multirow{2}{*}{\textbf {Clean}}\\
    {} & \multicolumn{1}{c}{\textbf{-5 dB}} & \multicolumn{1}{c}{\textbf{0 dB}} & \multicolumn{1}{c}{\textbf{5 dB}} & {} \\
    \hline
    Baseline & $36.5$ & $22.5$ & $14.0$ & $\textbf{6.7}$ \\
    \hline
    E0       & $33.6$ & $20.4$ & $13.3$ & $\textbf{6.7}$ \\
    E1       & $32.2$ & $19.8$  & $12.9$ & $\textbf{6.7}$ \\
    E2       & $31.9$ & $19.6$ & $\textbf{12.6}$ & $\textbf{6.7}$ \\
    E3       & $\textbf{31.8}$ & $\textbf{19.3}$  & $12.7$ & $6.8$ \\
    \hline
  \end{tabular}
\end{table}

The results on the simulated noisy Librispeech sets are shown in Tab.~\ref{tab:librispeech-noisecontext}. As can be seen, all enhancement frontends improve performance across SNRs. E0, which is the frontend model that does not use any noise context, provides relative WER improvements of 7.8\% at $-5$~dB, 9.3\% at $0$~dB and 4.5\% at $5$~dB. As one might expect, the gains are lower at higher SNRs. E3 corresponds to the model that makes use of noise context using the architecture described in Sec.~\ref{subsec:caconformer}. It uses the final formulation of the cross-attention conformer layer, with 2 cross-attention modules and the FiLM module. E3 improves WER by $\sim5$\% relative compared to E0. Compared to the baseline, it improves WER by 12.8\%, 14.1\% and 9.3\% at $-5$~dB, $0$~dB, and $5$~dB, respectively. We also present results using models that do not use the FiLM module (E2), and one without FiLM or the second cross attention (E1). E1 still outperforms E0 across conditions, but is worse than both E2 and E3. The results show that using noise context via the cross-attention conformer layers improves performance. E2 and E3 perform similarly, with E3 working marginally better at $-5$~dB and $0$~dB. As shown in the table, performance in clean conditions (test-clean subset of Librispeech) is not adversely affected by the enhancement frontend.

\begin{table}[tbh]
  \centering
  \caption{Results using noise context models on vendor collected test sets that was re-recorded and mixed with re-recorded noise. The baseline model uses no enhancement frontend. E0 corresponds to a frontend that does not use noise context. E3 is the model described in Sec.~\ref{sec:model} with 2 cross-attention and the FiLM modules. E2 is model without FiLM. E1 is a model without FiLM or the second cross-attention.}
  \label{tab:vendor-noisecontext}
  \begin{tabular}{lcccc}
    \hline
    \multirow{2}{*}{\textbf{Test set}} & \multicolumn{3}{c}{\textbf{SNR}} & \multirow{2}{*}{\textbf {Clean}}\\
    {} & \multicolumn{1}{c}{\textbf{0 dB}} & \multicolumn{1}{c}{\textbf{6 dB}} & \multicolumn{1}{c}{\textbf{12 dB}} & {}\\
    \hline
    Baseline & $46.3$ & $19.0$ & $8.6$ & $2.2$ \\
    \hline
    E0       & $42.5$ & $17.3$ & $7.6$ & $\textbf{2.1}$ \\
    E1       & $37.0$ & $15.1$  & $\textbf{6.4}$ & $2.2$ \\
    E2       & $35.1$ & $\textbf{14.2}$ & $6.5$ & $2.4$ \\
    E3       & $\textbf{34.6}$ & $15.0$  & $\textbf{6.4}$ & $2.2$ \\
    \hline
  \end{tabular}
\end{table}

Results on the more realistic vendor-collected dataset are shown in Tab.~\ref{tab:vendor-noisecontext}. The overall trends are quite similar to the results on the simulated Librispeech test sets, but overall, the enhancement frontend produces larger relative gains. For example, E0, which does not use any noise context, provides relative improvements of 8.2\% at 0~dB and 12\% at 12~dB. Similarly, E3 improves WER by 25.4\% at 0~dB compared to the baseline, and 25.1\% at 12~dB. In alignment with the results on the simulated sets, both E2 and E3 perform similar to each other. All models that use noise context using cross-attention layers outperform the baseline enhancement frontend that does not use any noise context. As with Librispeech, performance in clean conditions is not affected significantly by the enhancement frontend.

\begin{table}[tb]
  \centering
  \caption{Results using noise context models on multi-talker Librispeech test sets. The baseline model uses no enhancement frontend. E0 corresponds to a frontend that does not use noise context. E3 is the model described in Sec.~\ref{sec:model} with 2 cross-attention and the FiLM modules. E2 is model without FiLM. E1 is a model without FiLM or the second cross-attention.}
  \label{tab:librispeech-multitalker}
  \begin{tabular}{lccccc}
    \hline
    \multirow{2}{*}{\textbf{Test set}} & \multicolumn{3}{c}{\textbf{SNR}} & \multicolumn{2}{c}{\textbf{Clean}} \\
    {} & \multicolumn{1}{c}{\textbf{-5 dB}} & \multicolumn{1}{c}{\textbf{0 dB}} & \multicolumn{1}{c}{\textbf{5 dB}}  & \multicolumn{1}{c}{\textbf{test-clean}}  & \multicolumn{1}{c}{\textbf{test-other}} \\
    \hline
    Baseline & $69.2$ & $46.4$ & $29.3$ & $\textbf{6.7}$ & $\textbf{12.8}$ \\
    \hline
    E0       & $58.4$ & $39.2$ & $25.4$ & $\textbf{6.7}$ & $\textbf{12.8}$  \\
    E1       & $53.3$ & $35.2$  & $23.4$ & $\textbf{6.7}$ & $\textbf{12.8}$  \\
    E2       & $\textbf{50.3}$ & $33.8$ & $\textbf{22.8}$ & $\textbf{6.7}$ & $12.9$  \\
    E3       & $50.4$ & $\textbf{33.6}$  & $\textbf{22.8}$ & $6.8$ & $12.9$  \\
    \hline
  \end{tabular}
\vspace{-0.2in}
\end{table}

The results in the harder multi-talker conditions are shown in Tab.~\ref{tab:librispeech-multitalker}. Multi-talker conditions are especially hard for the ASR model, which is trained to recognize a single speaker and without any multi-talker data in its training set. As a result, the model has difficulties differentiating the target speaker and the competing talker, especially at lower SNRs. When using E0, which does not use any noise context, performance already improves over the baseline. But when using noise context, which provides some contextual information about the competing speaker, performance significantly improves. Compared to E0, E3 obtains between 10\% -- 14\% relative improvement. Compared to the baseline with no enhancement, E3 improves WER by 22\% -- 28\%. 

In summary, incorporating noise context helps improve performance in noisy conditions, without affecting performance when there is no noise. The cross-attention conformer layer proposed in this work provides an effective way to incorporate contextual information into the model.

\section{Conclusion}
\label{sec:concl}

In this work, we presented cross-attention conformers for incorporating contextual information. We do this by extending the conformer architecture, which combines convolution and attention modules, to also use contextual information as auxiliary inputs. The layer makes use of a cross-attention mechanism to combine the auxiliary features and the main features. 

Using noise context as the contextual information, we showed how the cross-attention layer can be used to improve performance of an enhancement frontend on an ASR task in the presence of noise. The model improves performance in both simulated and realistic conditions, as shown by the results on the simulated noisy Librispeech test sets, and the re-recorded vendor-collected datasets. Interestingly, the model also performs well in removing competing speech, in which case the noise context provides information about the background speech. The model effectively makes use of this information to enhance the noisy features, even when it is not trained in such conditions. Overall, the best model provides relative WER improvements of 9\% -- 28\% in low SNR settings, without affecting performance in clean conditions.

In future work, it will be interesting to consider other contextual signals, like multiple speakers~\cite{wang2021multiuser}, and input from other modalities like vision \cite{braga2020end}. Another interesting direction would be to extend the architecture to simultaneously handle multiple contextual signals.

\section{Acknowledgements}
\label{sec:ack}

We thank Alex Gruenstein, Alex Park, Mert Saglam, Nathan Howard, and Sankaran Panchapagesan for several useful discussions, feedback, and help with datasets.

\bibliographystyle{IEEEbib}
\bibliography{refs_subset}

\begin{thebibliography}{10}

\bibitem{PrabhavalkarRaoSainathLiEtAl17}
R.~Prabhavalkar, K.~Rao, T.~N. Sainath, B.~Li, L.~Johnson, and N.~Jaitly,
\newblock ``{A Comparison of Sequence-to-Sequence Models for Speech
  Recognition},''
\newblock in {\em Proc. Interspeech}, 2017.

\bibitem{BattenbergChenChildCoatesEtAl17}
E.~{Battenberg}, J.~{Chen}, R.~{Child}, A.~{Coates}, Y.~G.~Y. {Li}, H.~{Liu},
  S.~{Satheesh}, A.~{Sriram}, and Z.~{Zhu},
\newblock ``{Exploring Neural Transducers for End-to-end Speech Recognition},''
\newblock in {\em Proc. ASRU}, 2017.

\bibitem{HoriWatanabeZhangChan2017}
T.~Hori, S.~Watanabe, Y.~Zhang, and W.~Chan,
\newblock ``{Advances in Joint {CTC}-Attention Based End-to-End Speech
  Recognition with a Deep {CNN} Encoder and {RNN-LM}},''
\newblock in {\em Proc. Interspeech}, 2017.

\bibitem{li2021betterfaster}
B.~Li, A.~Gulati, J.~Yu, T.~N. Sainath, C.-. Chiu, A.~Narayanan, S.-Y. Chang,
  et~al.,
\newblock ``A better and faster end-to-end model for streaming {ASR},''
\newblock in {\em Proc. ICASSP}, 2021.

\bibitem{mirsamadi2017multi}
S.~Mirsamadi and J.~HL Hansen,
\newblock ``On multi-domain training and adaptation of end-to-end {RNN}
  acoustic models for distant speech recognition,''
\newblock in {\em Proc. Interspeech}, 2017.

\bibitem{hakkani2016multi}
D.~Hakkani-T{\"u}r, G.~T{\"u}r, A.~Celikyilmaz, Y-N. Chen, J.~Gao, L.~Deng, and
  Y.-Y. Wang,
\newblock ``Multi-domain joint semantic frame parsing using bi-directional
  {RNN-LSTM},''
\newblock in {\em Proc. Interspeech}, 2016.

\bibitem{NarayananMisraSimPundakEtAl18}
A.~{Narayanan}, A.~{Misra}, K.~C. {Sim}, G.~{Pundak}, A.~{Tripathi},
  M.~{Elfeky}, P.~{Haghani}, T.~{Strohman}, and M.~{Bacchiani},
\newblock ``{Toward Domain-Invariant Speech Recognition via Large Scale
  Training},''
\newblock in {\em Proc. of SLT}, 2018.

\bibitem{kim2017mtr}
C.~Kim, A.~Misra, K.~Chin, T.~Hughes, A.~Narayanan, T.~Sainath, and
  M.~Bacchiani,
\newblock ``Generation of large-scale simulated utterances in virtual rooms to
  train deep-neural networks for far-field speech recognition in {Google
  Home},''
\newblock in {\em Proc. Interspeech}, 2017.

\bibitem{park2019specaugment}
D.~S. Park, W.~Chan, Y.~Zhang, C.-C. Chiu, B.~Zoph, E.~D. Cubuk, and Q.~V. Le,
\newblock ``{SpecAugment}: {A} simple data augmentation method for automatic
  speech recognition,''
\newblock {\em Proc. Interspeech}, 2019.

\bibitem{medennikov2018investigation}
I.~Medennikov, Y.~Y. Khokhlov, A.~Romanenko, D.~Popov, N.~A. Tomashenko,
  I.~Sorokin, and A.~Zatvornitskiy,
\newblock ``An investigation of mixup training strategies for acoustic models
  in {ASR},''
\newblock in {\em Proc. Interspeech}, 2018.

\bibitem{barker2017thirdchime}
J.~Barker, R.~Marxer, E.~Vincent, and S.~Watanabe,
\newblock ``The third {CHiME} speech separation and recognition challenge:
  Analysis and outcomes,''
\newblock {\em Computer Speech \& Language}, vol. 46, pp. 605--626, 2017.

\bibitem{barker2018fifthchime}
J.~Barker, S.~Watanabe, E.~Vincent, and J.~Trmal,
\newblock ``The fifth {CHiME} speech separation and recognition challenge:
  Dataset, task and baselines,''
\newblock in {\em Proc. Interspeech}, 2018.

\bibitem{Li2014Review}
J.~Li, L.~Deng, Y.~Gong, and R.~Haeb-Umbach,
\newblock ``An overview of noise-robust automatic speech recognition,''
\newblock {\em IEEE/ACM Transactions on Audio, Speech, and Language
  Processing}, vol. 22, pp. 745--777, 2014.

\bibitem{chen2021continuous}
S.~Chen, Y.~Wu, Z.~Chen, J.~Wu, J.~Li, T.~Yoshioka, C.~Wang, S.~Liu, and
  M.~Zhou,
\newblock ``Continuous speech separation with conformer,''
\newblock in {\em Proc. ICASSP}, 2021.

\bibitem{luo2018tasnet}
Y.~Luo and N.~Mesgarani,
\newblock ``Tasnet: time-domain audio separation network for real-time,
  single-channel speech separation,''
\newblock in {\em Proc. ICASSP}, 2018.

\bibitem{Narayanan2013IRM}
A.~Narayanan and D.~L. Wang,
\newblock ``Ideal ratio mask estimation using deep neural networks for robust
  speech recognition,''
\newblock in {\em Proc. ICASSP}, 2013.

\bibitem{wang2021tunein}
J.~Wang, M.~W.~Y. Lam, D.~Su, and D.~Yu,
\newblock ``Tune-in: Training under negative environments with interference for
  attention networks simulating cocktail party effect,''
\newblock in {\em Proc. AAAI}, 2021.

\bibitem{tagliasacchi2020seanet}
M.~Tagliasacchi, Y.~Li, K.~Misiunas, and D.~Roblek,
\newblock ``Seanet: A multi-modal speech enhancement network,''
\newblock in {\em Proc. Interspeech}, 2020.

\bibitem{Seltzer2013DNNAurora4}
M.~L. Seltzer, D.~Yu, and Y.-Q. Wang,
\newblock ``An investigation of deep neural networks for noise robust speech
  recognition,''
\newblock in {\em Proc. ICASSP}, 2013.

\bibitem{huang2019hotwordcleaner}
Y.~A. Huang, T.~Z. Shabestary, and A.~Gruenstein,
\newblock ``Hotword cleaner: dual-microphone adaptive noise cancellation with
  deferred filter coefficients for robust keyword spotting,''
\newblock in {\em Proc. ICASSP}, 2019.

\bibitem{wang2019voicefilter}
Q.~Wang, H.~Muckenhirn, K.~Wilson, P.~Sridhar, Z.~Wu, J.~R. Hershey, R.~A.
  Saurous, R.~J. Weiss, Y.~Jia, and I.~L. Moreno,
\newblock ``{VoiceFilter}: Targeted voice separation by speaker-conditioned
  spectrogram masking,''
\newblock in {\em Proc. Interspeech}, 2019.

\bibitem{Wang2020}
Q~Wang, I.~L. Moreno, M.~Saglam, K.~Wilson, A.~Chiao, R.~L., Yanzhang He, Wei
  Li, Jason Pelecanos, Marily Nika, and Alexander Gruenstein,
\newblock ``{VoiceFilter-Lite}: Streaming targeted voice separation for
  on-device speech recognition,''
\newblock in {\em Proc. Interspeech}, 2020.

\bibitem{Narayanan2015djat}
A.~Narayanan and D.~L. Wang,
\newblock ``Improving robustness of deep neural network acoustic models via
  speech separation and joint adaptive training,''
\newblock {\em IEEE/ACM Transactions on Audio, Speech, and Language
  Processing}, vol. 23, pp. 92--101, 2015.

\bibitem{vaswani2017attention}
A.~Vaswani, N.~Shazeer, N.~Parmar, J.~Uszkoreit, L.~Jones, A.~N Gomez,
  L.~Kaiser, and I.~Polosukhin,
\newblock ``Attention is all you need,''
\newblock in {\em Advances in Neural Information Processing Systems}, 2017.

\bibitem{gulati2020conformer}
A.~Gulati, J.~Qin, C.-C. Chiu, N.~Parmar, Y.~Zhang, J.~Yu, W.~Han, S.~Wang,
  Z.~Zhang, Y.~Wu, et~al.,
\newblock ``Conformer: Convolution-augmented transformer for speech
  recognition,''
\newblock {\em Proc. Interspeech}, 2020.

\bibitem{yeh2019transformer}
C.-F. Yeh, J.~Mahadeokar, K.~Kalgaonkar, Y.~Wang, D.~Le, M.~Jain, K.~Schubert,
  C.~Fuegen, and M.~L. Seltzer,
\newblock ``Transformer-transducer: End-to-end speech recognition with
  self-attention,''
\newblock {\em arXiv preprint arXiv:1910.12977}, 2019.

\bibitem{karita2019comparative}
S.~Karita, N.~Chen, T.~Hayashi, T.~Hori, H.~Inaguma, Z.~Jiang, M.~Someki,
  N.~E.~Y. Soplin, R.~Yamamoto, X.~Wang, et~al.,
\newblock ``A comparative study on transformer vs rnn in speech applications,''
\newblock in {\em Proc. IEEE ASRU}, 2019.

\bibitem{subakan2021attention}
C.~Subakan, M.~Ravanelli, S.~Cornell, M.~Bronzi, and J.~Zhong,
\newblock ``Attention is all you need in speech separation,''
\newblock in {\em Prof. ICASSP}, 2021.

\bibitem{graves12rnnt}
A.~{Graves},
\newblock ``{Sequence Transduction with Recurrent Neural Networks},''
\newblock {\em arXiv preprint arXiv:1211.3711}, 2012.

\bibitem{chan2016listen}
W.~Chan, N.~Jaitly, Q.~Le, and O.~Vinyals,
\newblock ``Listen, attend and spell: A neural network for large vocabulary
  conversational speech recognition,''
\newblock in {\em Proc. ICASSP}, 2016.

\bibitem{li2020parallel}
W.~Li, J.~Qin, C.-C. Chiu, R~Pang, and Y.~He,
\newblock ``Parallel rescoring with transformer for streaming on-device speech
  recognition,''
\newblock in {\em Proc. Interspeech}, 2020.

\bibitem{ding2020textual}
S.~Ding, Y.~Jia, K.~Hu, and Q.~Wang,
\newblock ``Textual echo cancellation,''
\newblock {\em arXiv preprint arXiv:2008.06006}, 2020.

\bibitem{sainath2020streaming}
T.~N. Sainath, Y.~He, B.~Li, A.~Narayanan, R.~Pang, A.~Bruguier, S.-Y. Chang,
  W.~Li, R.~Alvarez, Z.~Chen, et~al.,
\newblock ``A streaming on-device end-to-end model surpassing server-side
  conventional model quality and latency,''
\newblock in {\em Proc. ICASSP}, 2020.

\bibitem{perez2018film}
Strub F. de Vries H. Dumoulin~V. Perez, E. and A.~Courville,
\newblock ``{FiLM}: Visual reasoning with a general conditioning layer,''
\newblock in {\em Proceedings of the AAAI Conference on Artificial
  Intelligence}, 2018.

\bibitem{DaiYangYangCarbonnelEtAl19}
Z.~{Dai}, Z.~{Yang}, Y.~{Yang}, J.~{Carbonell}, Q.~V. {Le}, and
  R.~{Salakhutdinov},
\newblock ``{Transformer-XL: Attentive Language Models Beyond a Fixed-Length
  Context},''
\newblock {\em arXiv preprint arXiv:1901.02860}, 2019.

\bibitem{wang2020transformer}
Y.~Wang, A.~Mohamed, D.~Le, C.~Liu, A.~Xiao, J.~Mahadeokar, H.~Huang,
  A.~Tjandra, X.~Zhang, F.~Zhang, et~al.,
\newblock ``Transformer-based acoustic modeling for hybrid speech
  recognition,''
\newblock in {\em Prof. ICASSP}, 2020.

\bibitem{bahdanau2014neural}
D.~Bahdanau, K.~Cho, and Y.~Bengio,
\newblock ``Neural machine translation by jointly learning to align and
  translate,''
\newblock {\em arXiv preprint arXiv:1409.0473}, 2014.

\bibitem{panayotov2015librispeech}
V.~Panayotov, G.~Chen, D.~Povey, and S.~Khudanpur,
\newblock ``Librispeech: An {ASR} corpus based on public domain audio books,''
\newblock in {\em Proc. ICASSP}, 2015.

\bibitem{Narayanan2014Joint}
A.~Narayanan and D.~L. Wang,
\newblock ``Joint noise adaptive training for robust automatic speech
  recognition,''
\newblock in {\em Proc. ICASSP}, 2014.

\bibitem{abadi2016tensorflow}
M.~Abadi, P.~Barham, J.~Chen, Z.~Chen, A.~Davis, J.~Dean, M.~Devin,
  S.~Ghemawat, G.~Irving, M.~Isard, et~al.,
\newblock ``Tensorflow: A system for large-scale machine learning.,''
\newblock in {\em OSDI}, 2016, vol.~16, pp. 265--283.

\bibitem{ShenNguyenWuChenEtAl19}
J.~{Shen}, P.~{Nguyen}, Y.~{Wu}, Z.~{Chen}, M.~X. {Chen}, Y.~{Jia},
  A.~{Kannan}, T.~{Sainath}, Y.~{Cao}, C.-C. {Chiu}, et~al.,
\newblock ``{Lingvo: A Modular and Scalable Framework for Sequence-to-Sequence
  Modeling},''
\newblock {\em arXiv preprint arXiv:1902.08295}, 2019.

\bibitem{jouppi2017datacenter}
N.~P. Jouppi, C.~Young, N.~Patil, D.~Patterson, G.~Agrawal, R.~Bajwa, S.~Bates,
  S.~Bhatia, N.~Boden, A.~Borchers, et~al.,
\newblock ``In-datacenter performance analysis of a tensor processing unit,''
\newblock in {\em Computer Architecture (ISCA), 2017 ACM/IEEE 44th Annual
  International Symposium on}. IEEE, 2017, pp. 1--12.

\bibitem{kingma2014adam}
D.~P. Kingma and J.~Ba,
\newblock ``Adam: A method for stochastic optimization,''
\newblock {\em arXiv preprint arXiv:1412.6980}, 2014.

\bibitem{wang2021multiuser}
R.~Rikhye, Q.~Wang, Q.~Liang, Y.~He, and I.~McGraw,
\newblock ``Multi-user voicefilter-lite via attentive speaker embedding,''
\newblock in {\em Under review, IEEE ASRU}, 2021.

\bibitem{braga2020end}
O.~Braga, T.~Makino, O.~Siohan, and H.~Liao,
\newblock ``End-to-end multi-person audio/visual automatic speech
  recognition,''
\newblock in {\em Proc. ICASSP}, 2020.

\end{thebibliography}

\end{document}